# Metal-insulator transition and superconductivity induced by Rh doping in binary Ru Pnictides RuPn (Pn = P, As and Sb)

D. Hirai[1*], T. Takayama[1,2], D. Hashizume[3] and H. Takagi[1,2,3]

[1] Department of Advanced Materials, University of Tokyo, Kashiwa 277-8561, Japan
[2] JST-TRIP, Kashiwa 277-8561, Japan
[3] RIKEN Advanced Science Institute, Wako 351-0198, Japan

## Abstract

Binary ruthenium pnictides, RuP and RuAs, with an orthorhombic MnP structure, were found to show a metal to a non-magnetic insulator transition at $T_{MI}$ = 270 K and 200 K, respectively. In the metallic region above $T_{MI}$, a structural phase transition, accompanied by a weak anomaly in the resistivity and the magnetic susceptibility, indicative of a pseudo-gap formation, was identified at $T_s$ = 330 K and 280 K, respectively. These two transitions were suppressed by substituting Ru with Rh. We found superconductivity with a maximum $T_c$ = 3.7 K and $T_c$ =1.8 K in a narrow composition range around the critical point for the pseudo-gap phase, Rh content $x_c$ = 0.45 and $x_c$ = 0.25 for $Ru_{1-x}Rh_xP$ and $Ru_{1-x}Rh_xAs$, respectively, which may provide us with a novel non-magnetic route to superconductivity at a quantum critical point.

The relationship between superconductivity and other collective electronic states has been a long-standing enigma in condensed matter physics. In a variety of systems with distinct chemical characters, including cuprates [1], heavy fermions [2], organics [3] and more recently Fe pnictides [4], superconductivity was found in a narrow region near the critical boarder to magnetism as a function of pressure and doping. Superconductivity has also been observed at a critical border to classes of electronic orderings other than magnetic order, including charge ordering [5-7] and charge density wave [8, 9], although $T_c$ remains relatively low.

Stimulated by the discovery of Fe pnictide superconductors, we have been exploring Ru pnictides. 4$d$ Ru has the same $d$-electron number as 3$d$ Fe and is in general less magnetic. The simplest Ru pnictides are binary. A series of binary compounds RuPn (Pn = P, As and Sb) have been reported to crystallize in a MnP-type orthorhombic structure (space group *Pnma*) [10-12]. In this crystal structure (see the inset to Fig. 1), RuPn$_6$ octahedra form a face sharing chain along the $a$-axis. The chains are connected by the edges and Ru forms a distorted triangular lattice within the *bc*-plane.

We discovered two sequential phase transitions in RuP and RuAs: a first order transition from a metal to a non-magnetic insulator at low temperature ($T_{MI}$ = 270 K for RuP and 200 K for RuAs) and a weak transition to a pseudo-gap phase accompanied with the superstructure formation at high temperature ($T_s$ = 330 K for RuP and 280 K for RuAs). We could have suppressed those two transitions by Rh doping for Ru and found superconductivity at the critical point for the "pseudo-gap phase". Although the microscopic origin of the transitions remains yet to be clarified, the discovery should provide a new playground for superconductivity at a non-magnetic critical point. In this letter, we present the transport, magnetic, thermal and structural properties of (Ru, Rh)P and (Ru, Rh)As with emphasis on the discovery of two phase transitions and superconductivity at a critical point, and discuss the possible origin of the phase transitions.

Polycrystalline samples of RuPn (Pn = P, As and Sb) and Rh doped samples (Ru, Rh)P and (Ru, Rh)As were prepared by a conventional solid state reaction. A mixture of Ru metal, Rh metal and pnictogen elements was sintered in an evacuated quartz tube initially at 550 ºC for 10 h and then at 1050 ºC for (Ru, Rh)P, 950 ºC for (Ru, Rh)As and 900 ºC for RuSb for 48 hours. An excess of pnictogen elements was added to compensate for the loss due to volatilization. The sintered pellet was reground, repelletized and sintered again for 72 hours. The phase purity was checked by powder x-ray diffractions using Cu-$K\alpha$ radiation. Magnetic, transport and thermal measurements were conducted by a SQUID magnetometer and a Physical Property Measurement System (PPMS: Quantum Design). Electric resistivity above 350 K was measured separately by a four-probe method in a furnace with flowing $N_2$ gas. Very small single crystals of RuP used for the structural analysis were grown out of Sn flux.

All the three pnictides, RuP, RuAs and RuSb were found to be poorly metallic at room temperature with almost temperature independent resistivity $\rho(T)$ of ~ 1 mΩcm, as shown in Fig. 2. On cooling, a metal-insulator transition was clearly observed for RuP and RuAs at $T_{MI}$ = 270 K and 200 K, shown in Fig. 2(a) and, below the $T_{MI}$, $\rho(T)$ shows an insulating behavior. The presence of tiny but clear hysteresis around the $T_{MI}$ indicates that the metal insulator transitions are of a first order. In the case of RuAs, we observe a much broader transition than in RuP, which we believe represents the presence of some inhomogeneity in the RuAs sample. RuSb was found to be metallic down to the lowest temperature measured.

The magnetic susceptibility $\chi(T)$ in the metallic phase above $T_{MI}$ is less than $10^{-4}$ emu/mol, which may be ascribed to the Pauli paramagnetic susceptibility of a metal with a moderate density of states. At $T_{MI}$ for RuP and RuAs, $\chi(T)$ shows an almost discontinuous drop to a negative value with hysteresis, which is comparable to the expected core diamagnetism [13]. This suggests that the low temperature insulating state is non-magnetic. In fact, the preliminary μSR experiment on RuAs [14] was consistent with the non-magnetic ground state. The systematic suppression of a metal-insulator transition on going from P, As to Sb very likely reflects the increased band width due to the enhanced $p$-$d$ hybridization, but the increase of magnetic susceptibility from P to Sb in the metallic state may not allow such naive interpretation at least in its simplest form.

Closely inspecting the poorly metallic state of RuP and RuAs above $T_{MI}$, we notice an additional anomaly at $T_s$ = 330 K and 280 K, respectively. As shown in the inset to Fig. 2, at $T_s$, $\rho(T)$ shows a minimum and $\chi(T)$ shows a maximum. It appears that the anomaly at $T_s$ represents a precursor to the metal to non-magnetic insulator transition in that $\rho(T)$ increases and $\chi(T)$ decreases below $T_s$.

From $\rho(T)$ and $\chi(T)$ data alone, it is not clear whether or not $T_s$ represents a well-defined phase transition. However, the structural analysis on a RuP single crystal indicates clearly that it is a phase transition. The crystal structure of RuP at 400 K (above $T_s$) was refined well with an orthorhombic *Pnma* space group as reported in [10]. On cooling, superlattice spots $h\ k/4\ l/4$ appears just below $T_s = 330$ K, indicating the 4-fold structural modulation within the *bc* plane along [011] direction. By lowering the temperature further, additional spots indicative of tripling of the *a*-axis, the chain direction, emerge at $T_{MIT} = 270$ K. The crystal structures below $T_s$ and below $T_{MI}$ remain yet to be clarified. In the RuAs polycrystalline powder, we observed the super-lattice peaks in the powder pattern at $T_s$ and $T_{MI}$, analogous to those observed for the RuP single crystal.

Our preliminary band structure calculation indicated the presence of complicated and multiple Fermi surfaces. A nesting driven charge density wave in its simplest form is therefore highly unlikely to be able to describe the insulating ground state with the whole Fermi surface gapped, and a more elaborate picture, such as local spin dimer formation associated with orbital ordering, should be invoked. A metal to non-magnetic insulator transition in three-dimensional complex transition metal oxides has been observed, for example, in Magnéli phase vanadium and titanium oxides [15], $Tl_2Ru_2O_7$ [16], $CuIr_2S_4$ [17], $MgTi_2O_4$ [18] and $LiRh_2O_4$ [19]. In all these compounds, orbital ordering is believed to play a key role in realizing the non-magnetic, spin singlet ground state. Interestingly, in $LiRh_2O_4$, the orbital ordering, with weak $\rho(T)$ and $\chi(T)$ anomalies similar to those observed in Ru pnictides, occurs at a higher temperature than the first order metal-insulator transition and gives rise to a reduced dimensionality of the itinerant electrons, which acts as a precursor to the metal-nonmagnetic insulator transition [19]. To discuss further along this line, we should disclose the distortion pattern in the two low temperature ordered states.

Inspired by the close link between electronic order and superconductivity recognized in a variety of systems, we have attempted to suppress the two transitions in RuP and RuAs by doping. We found that Rh doping for Ru systematically suppresses the two transitions. As seen from $\rho(T)$ and $\chi(T)$ shown in Fig. 3, upon Rh-doping, the first order transition at $T_{MI}$ is rapidly suppressed and is absent already at 10% doping level for both RuP and RuAs. The transition at $T_s$ appears to be much more robust against doping than the metal-insulator transition. Even with more than 10% doping, we see a broad peak in $\chi(T)$ and a minimum in $\rho(T)$ representing $T_s$. Below $T_s$, an anomalous and poorly metallic state is realized. First of all, $\rho(T)$ shows a very weak increase on cooling but appears to approach a finite value. $\chi(T)$ shows a pronounced decrease on cooling sometimes even to a diamagnetic regime, which we in this letter call the poorly metallic phase below $T_s$ as a "pseudo-gap" phase. This is again suggestive of the transition at $T_s$ being a precursor to the non-magnetic insulator phase observed in the undoped compounds.

Eventually the pseudo-gap transition disappears at $x_c = 0.45$ for RuP and at $x_c = 0.25$ for RuAs as clearly seen in Fig. 3. In support of the presence of a well-defined critical point, the very clear anomaly in the doping dependence of Debye temperature $\Theta_D$ and the electronic specific coefficient $\gamma$ estimated from the specific heat was observed at $x_c$, indicative of the presence of phase transition involving both electrons and lattices.

We discovered superconductivity at the critical point for the pseudo-gap phase. As shown in Fig. 4(a), zero resistance and full diamagnetic shielding, indicative of a superconducting transition, are observed below $T_c = 3.7$ K and 1.8 K for the samples with the critical Rh content $x_c$, $Ru_{0.55}Rh_{0.45}P$ and $Ru_{0.75}Rh_{0.25}As$, respectively. The electronic specific heat $C_e(T)$ of those two samples were estimated by subtracting the normal state $C_N(T)$ under 9 T magnetic field, which is well above the critical field $\mu_0H_{c2}(0)$, and adding the $\gamma T$ term with $\gamma$ obtained from the extrapolation of $C_N(T)/T$ to $T = 0$. The electronic specific heat coefficient $\gamma$ was estimated as 1.3 mJ/mol·K$^2$ for $Ru_{0.55}Rh_{0.45}P$ and 3.0 mJ/mol·K$^2$ for $Ru_{0.75}Rh_{0.25}As$, which is quite moderate as a $4d$ inter-metallic compound. $C_e(T)$ shows a large jump at $T_c$ both for $Ru_{0.55}Rh_{0.45}P$ and $Ru_{0.75}Rh_{0.25}As$, evidencing for the bulk superconductivity. The rapid decrease of $C_e(T)/T$ below $T_c$ in $Ru_{0.55}Rh_{0.45}P$ suggests a gapful superconductivity, which is very likely $s$-wave. The slow decrease of $C_e(T)/T$ in $Ru_{0.75}Rh_{0.25}As$ at a glance appears to imply a gapless superconductivity, but considering the pronounced inhomogeneity in RuAs system, we suspect that it reflects a distribution of inhomogeneous gap rather than gap node(s).

As seen from the specific heat data $C(T)$ and magnetization data $\chi(T)$ for the samples with different doping levels shown in Fig. 4(b) and (c), superconductivity was observed in a limited region around the critical point $x_c$ and transition temperature $T_c$ peaked at $x_c$ both for $Ru_{1-x}Rh_xP$ and $Ru_{1-x}Rh_xAs$. The interplay between the criticality and superconductivity in doped RuP and RuAs can be illustrated visually as a phase diagram such as that shown in Fig. 1. The rapid collapse of the non-magnetic insulating phase upon doping may suggest that the accommodation of an integer number of electrons is an important ingredient for the emergence of a non-magnetic insulator phase below $T_{MI}$. On the other hand, the insensitiveness of $T_s$ and systematic suppression of the pseudo-gap behavior upon doping might mean a local character of the phase transition. The presence of superconducting dome centered at the critical point clearly indicates the link between the criticality to the ordering below $T_s$ and superconductivity.

Comparing (Ru,Rh)P and (Ru, Rh)As, it is clear that the $T_s$ ordering is suppressed more readily for (Ru, Rh)As in that $T_s$ for undoped compound and $x_c$ are much lower for RuAs than RuP. Possibly reflecting this, the "optimum" $T_c$ is higher for (Ru,Rh)P ($T_c$ = 3.7 K) than (Ru, Rh)As ($T_c$ = 1.8 K). It might be interesting to infer here that $Ru_{0.55}Rh_{0.45}P$ has smaller electronic specific heat coefficient ($\gamma$ ~1.3 mJ/mol·K$^2$) than that of $Ru_{0.75}Rh_{0.25}As$ ($\gamma$ ~ 3.0 mJ/mol·K$^2$). We argue that such an anti-correlation between DOS and $T_c$, opposite to what is predicted from BCS theory, might imply the vital role of the energy scale of criticality. The $T_s$ ordering appears to be suppressed completely for RuSb but superconductivity with a lower $T_c$ than (Ru,Rh)P and (Ru, Rh)As, $T_c$ = 1.2 K, was still observed, as seen in Fig. 4(a). This might suggest that RuSb is located not far away from the hidden critical point.

In conclusion, we found two sequential transitions in binary pnictides RuP and RuAs: a high temperature transition to a pseudo-gap phase at $T_s$ and a low temperature metal to non-magnetic insulator at $T_{MI}$. To clarify the physics behind those two transitions, the refinement of the lattice distortion pattern below $T_s$ and $T_{MI}$ should have a high priority. Rh doping was found to suppress those two transitions. In a narrow doping region around a critical point for the pseudo-gap phase, superconductivity was discovered with maximum $T_c$ of 3.7 K for $Ru_{0.55}Rh_{0.45}P$ and 1.8 K for $Ru_{0.75}Rh_{0.25}As$, giving rise to a novel playground for the superconductivity at a critical point. We emphasize here that the critical point here is neither antiferromagnetic nor ferromagnetic as is usually the case in widely discussed superconductivity at a critical point.

We thank A. Mackenzie, P. Radaelli, T. Mizokawa, Y. Katsura D. Nishio-Hamane and A. Yamamoto for stimulating discussions. This work was partly supported by Grant-in-Aid for Scientific Research (S) (Grant No. 19104008), Grant-in-Aid for Scientific Research on Priority Areas (Grant No. 19052008).

**Figures**

**Fig. 1** (color online) Electronic phase diagrams of $Ru_{1-x}Rh_xP$, $Ru_{1-x}Rh_xAs$ and RuSb as a function of Rh doping. Filled squares and triangles in $Ru_{1-x}Rh_xP$ and $Ru_{1-x}Rh_xAs$ correspond to the transition temperature to the pseudo-gap phase $T_s$ determined from the minima in $\rho(T)$ curves. Filled and open circles in $Ru_{1-x}Rh_xP$ represent the superconducting transition temperature $T_c$ determined from the magnetization and the specific heat measurements, respectively. Open triangles in $Ru_{1-x}Rh_xAs$ indicate $T_c$ determined from the heat capacity data. The inset shows the crystal structure of RuPn (Pn = P, As and Sb).

**Fig. 2** (color online) Temperature dependences of (a) resistivity $\rho(T)$ and (b) dc magnetic susceptibility $\chi(T)$ for RuP, RuAs and RuSb. Magnetic susceptibility was measured under applied field of 1 T. Open and filled arrows indicate the metal to non-magnetic insulator transitions and high-temperature structural transitions in RuP and RuAs, respectively. The insets show the $\rho(T)$ and $\chi(T)$ anomaly associated with phase transition at $T_s = 330$ K in RuP.

**Fig. 3** (color online) Temperature dependent (a, c) resistivity $\rho(T)$ and (b, d) dc magnetic susceptibility $\chi(T)$ for $Ru_{1-x}Rh_xPn$ (Pn = P and As). Magnetic susceptibility was measured under applied field of 1 T. The arrows in the insets of (a) and (c) indicate the minima of $\rho(T)$ curve, defined as the pseudo-gap transition temperature $T_s$.

**Fig. 4** (color online) Superconducting transitions observed in Ru pnictides. (a) Temperature dependent resistivity $\rho(T)$ of $Ru_{0.55}Rh_{0.45}P$, $Ru_{0.75}Rh_{0.25}As$ and RuSb. (b) Electronic specific heat divided by temperature $C_e/T$ for $Ru_{0.55}Rh_{0.45}P$ and $Ru_{1-x}Rh_xAs$ ($x$ = 0.15, 0.25, 0.35 and 0.45). (c) Dc magnetization data at low temperatures under applied magnetic field of 20 Oe for $Ru_{1-x}Rh_xP$ ($x$ = 0.35, 0.40, 0.45 and 0.50).

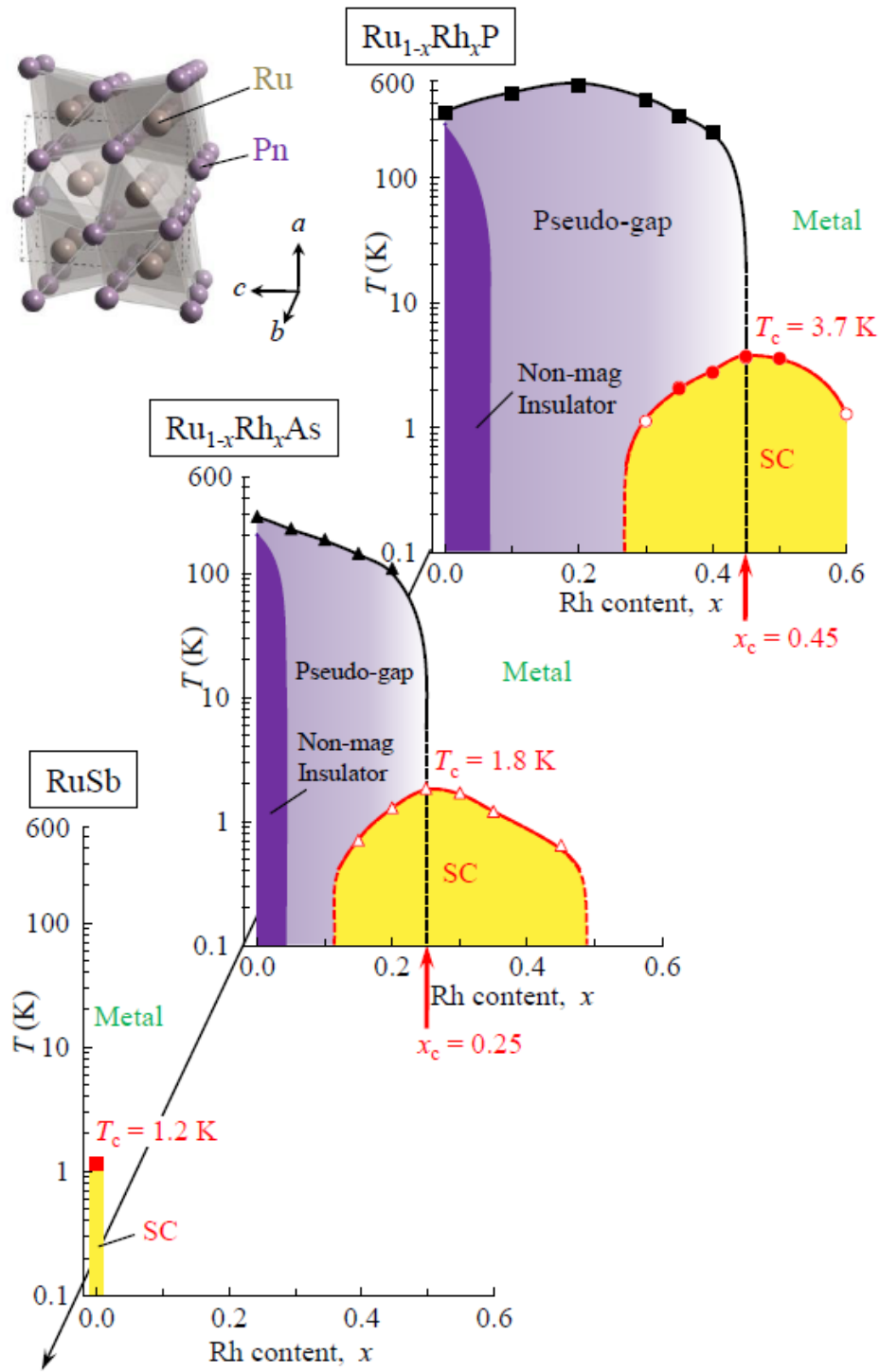

**Fig. 1**

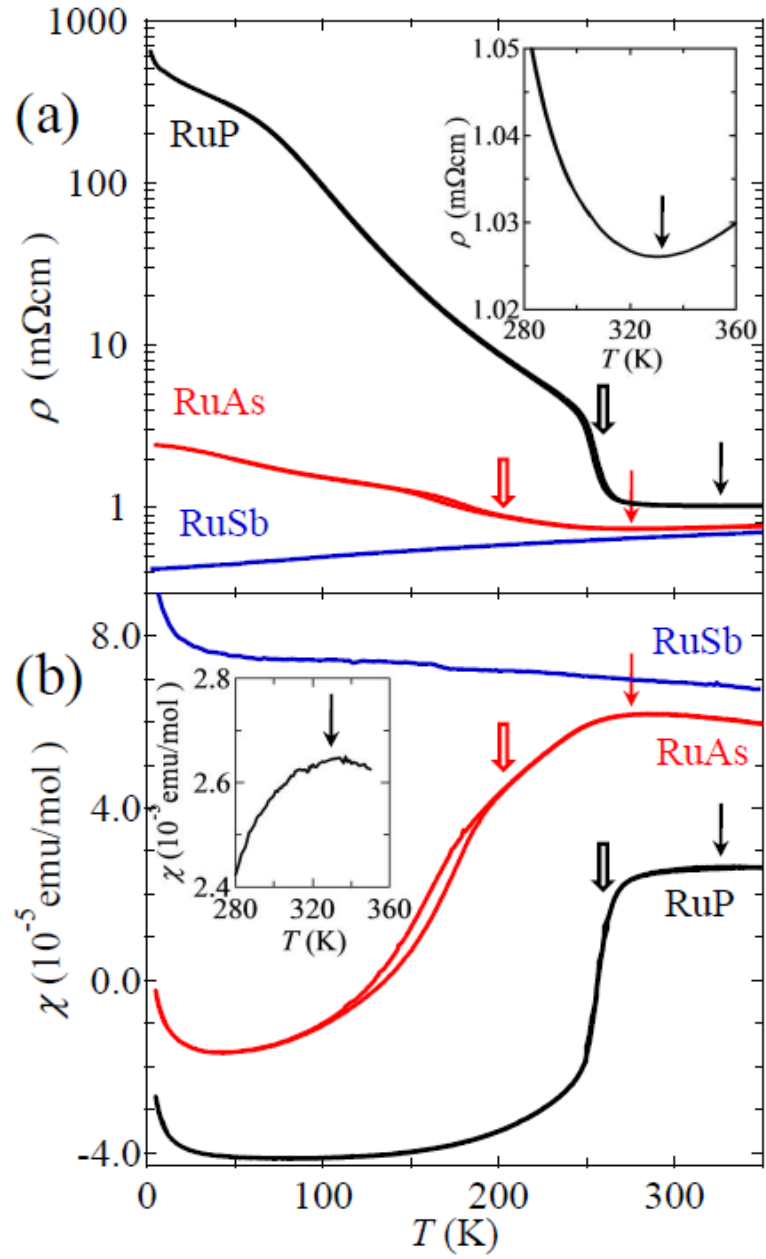

**Fig. 2**

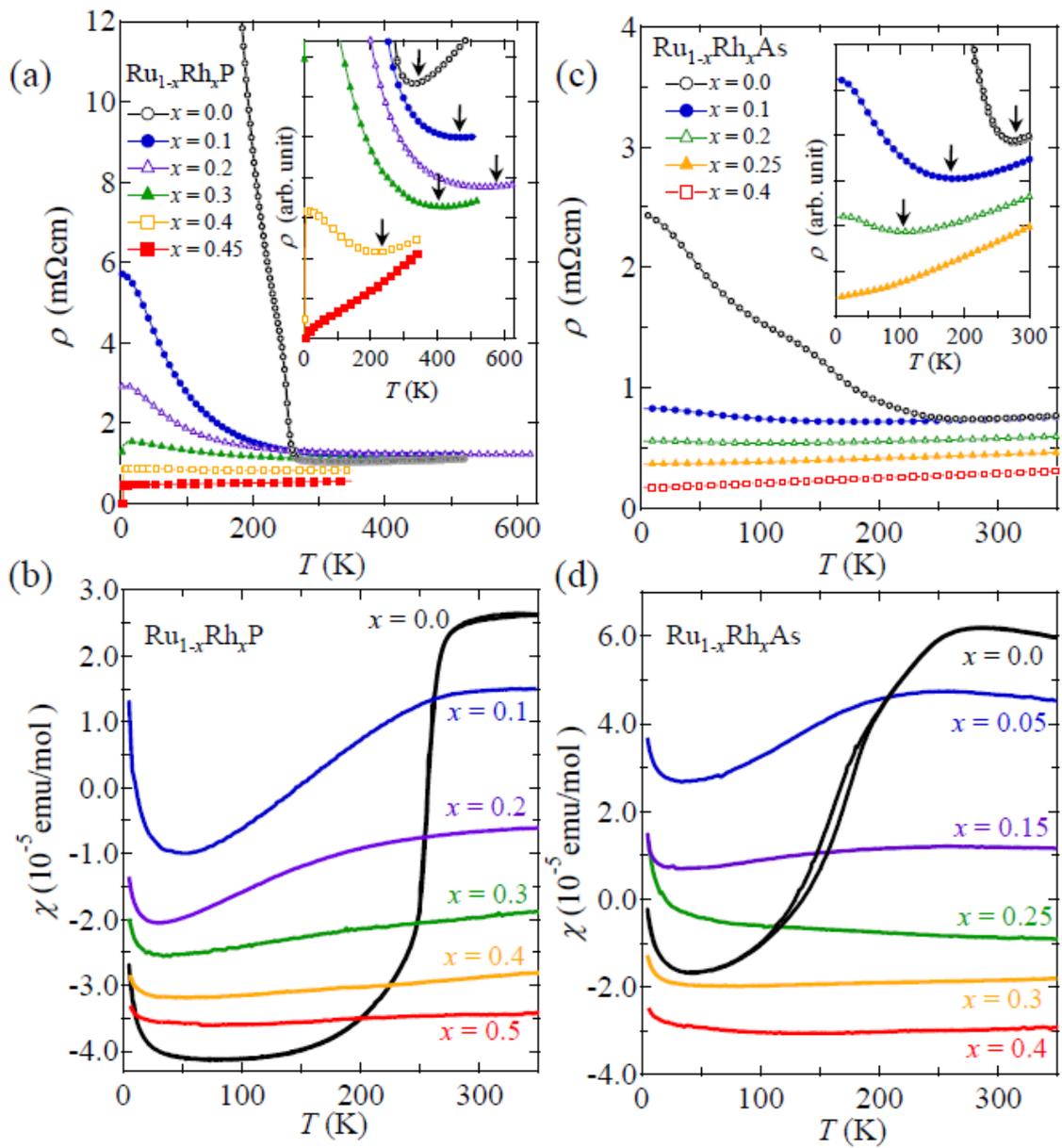

Fig. 3

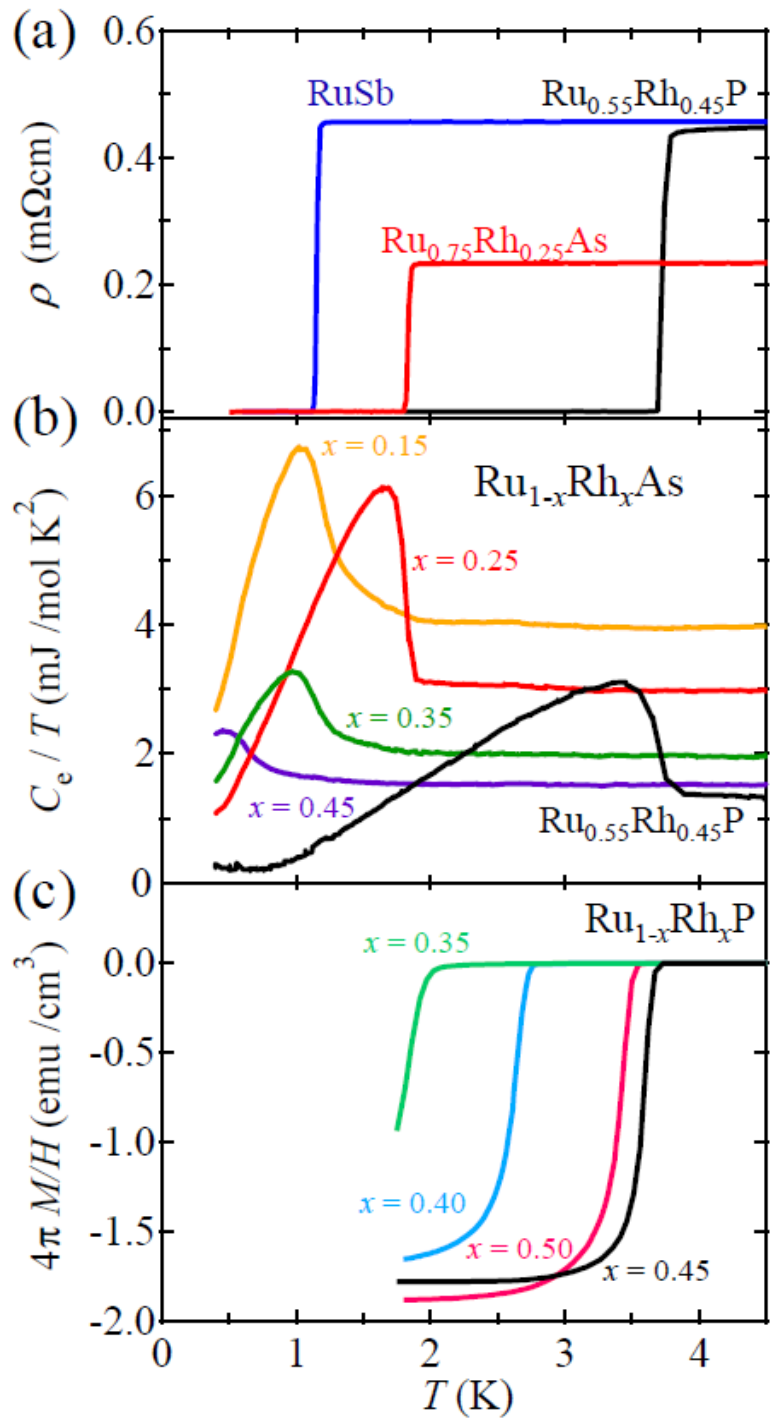

Fig. 4